\documentclass[twocolumn,pra]{revtex4-1}
\usepackage[latin9]{inputenc}
\usepackage{amsmath}
\usepackage{amssymb}
\usepackage{graphicx}

\makeatletter

\@ifundefined{textcolor}{}
{%
 \definecolor{BLACK}{gray}{0}
 \definecolor{WHITE}{gray}{1}
 \definecolor{RED}{rgb}{1,0,0}
 \definecolor{GREEN}{rgb}{0,1,0}
 \definecolor{BLUE}{rgb}{0,0,1}
 \definecolor{CYAN}{cmyk}{1,0,0,0}
 \definecolor{MAGENTA}{cmyk}{0,1,0,0}
 \definecolor{YELLOW}{cmyk}{0,0,1,0}
 }

\usepackage{dcolumn}\usepackage{bm}\usepackage{colordvi}

\makeatother

\begin{document}

\title{State tomography via weak measurements}

\author{Shengjun Wu\footnote{Correspondence to shengjun@ustc.edu.cn}}

\affiliation{Hefei National Laboratory for Physical Sciences at Microscale and
Department of Modern Physics, University of Science and Technology
of China, Hefei, Anhui 230026, China}

\begin{abstract}
Recent work has revealed that the wave function of a pure state can be measured directly and that complementary knowledge of a quantum system can be obtained simultaneously by weak measurements. However, the original scheme applies only to pure states, and it is not efficient because most of the data are discarded by post-selection. Here, we propose tomography schemes for pure states and for mixed states via weak measurements, and our schemes are more efficient because we do not discard any data. Furthermore, we demonstrate that any matrix element of a general state can be directly read from an appropriate weak measurement. The density matrix (with all of its elements) represents all that is directly accessible from a general measurement.
\end{abstract}

\maketitle

For a group of blindfolded observers, who can each only touch one part of the elephant, to construct an accurate representation of an elephant,
the group must combine information about different parts of the elephant (FIG 1a). In quantum mechanics, to have a complete description of a quantum system, in particular, a complete knowledge of a quantum state, we must combine information about complementary aspects (FIG 1b).

An ideal (strong) quantum measurement of a certain observable only gives the probabilities of obtaining the eigenvalues of the observable, and the statistical results of the measurements reflect the diagonal terms of the density matrix in the eigenbasis of the observable. Measurements of different (complementary) observables provide the diagonal terms of the density state for different bases. If we perform measurements for a sufficient number of bases, we can reconstruct the density state, which is the idea of state tomography.

An important difference exists between the tomography of a quantum state and its classical counterpart, e.g., the description of an elephant by blindfolded observers. In this classical example, the description (of the elephant) by each man is not exclusive, and the different aspects of perception can be simply combined to produce an overall description of the elephant (FIG 1a). However, quantum physics forbids simultaneous knowledge of complementary observables because precise knowledge of a certain aspect necessarily implies uncertainty for the complementary aspects \cite{Bohr28}.
Therefore, one cannot simultaneously perform ideal measurements of complementary observables in quantum physics (FIG 2a). To determine the quantum state of a system, different measurement setups, each for a particular observable, are required. To determine the general state of a $d$-dimensional quantum system, at least $d+1$ different experimental setups are needed. Each setup must be devoted to the measurement of one of the complementary observables.

\begin{figure}
\centering
\includegraphics[width=8cm]{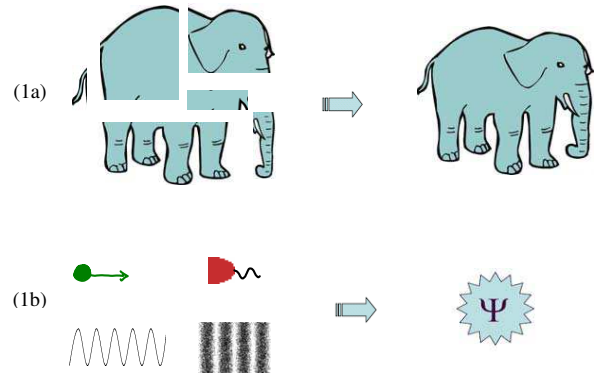}
\caption{(a) Compatible aspects are combined in classical physics, (b) while complementary aspects are combined in quantum physics.}
\label{fig1}
\end{figure}

An alternative state tomography approach is possible. Instead of obtaining maximum information of a particular observable by an ideal measurement, we can perform a weak measurement \cite{AAV}. The state of the quantum system is only slightly changed during its weak interaction with the measuring device; therefore, weak measurements of a set of complementary observables can be performed simultaneously \cite{Kocsis12} (FIG 2b). Thus, fewer experimental setups are needed for state tomography via weak measurements \cite{LSPSB2011}.

\begin{figure}
\centering
\includegraphics[width=8cm]{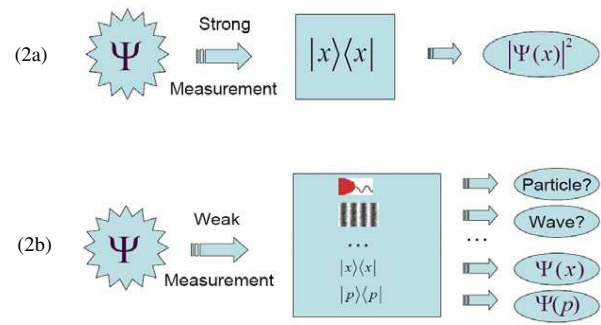}
\caption{(a) A strong measurement only reveals a certain aspect of the quantum state, (b) while a weak measurement enables us to perceive (measure) incompatible aspects of the quantum state simultaneously.}
\label{fig2}
\end{figure}

Weak measurements have been well described in the literature.  Since they were first introduced by Aharonov, Albert and Vaidman \cite{AAV}, weak measurements have been realised in experiments \cite{Ritchie,Pryde05,HK2008}, and have provided new insights into the study of paradoxes and fundamental problems in quantum theory \cite{ABPRT02,Wiseman07,Mir07,WJ08,Johansen}.  Weak measurement has also been used as a practical tool for amplifying weak signals and signal-to-noise ratios \cite{QSHE, Dixon, Starling, BS, FXS2011, ZRG2011, AP2007, WZ2012}. Signal amplification via weak measurement usually involves a pre-selection and a post-selection that are nearly orthogonal, and the original formalism of weak measurement \cite{AAV,Jozsa} is not sufficient for the phenomena in this regime because it only retains the first-order terms of the interaction strength \cite{Duck89,SDWJH,Geszti}. Extensions to the case of general pre-selection are given in \cite{Wiseman02,WM09}. A framework that retains the high-order terms for general pre-selection and post-selection is given in \cite{WL2011}, and specific cases are further studied in \cite{zhu2011,NNF2012,PWC2012}. A nonperturbative theory of weak pre- and post-selected measurements is given in \cite{KAN2012}.
The general results for the weak measurement of a pair of complementary observables are provided in \cite{WZ2012}. Related concepts, such as the contextual values \cite{DAJ10} and modular values \cite{KV}, were also introduced.
Weak measurements can also be used for state tomography. Lundeen et al. \cite{LSPSB2011} directly measured the wave function of a pure state by a weak measurement followed by a post-selection. The scheme was extended to the mixed state by sequential weak measurements of pairs or triple products of complementary observables \cite{LB12}.
The results of the weak measurements could also be interpreted in terms of the Kirkwood-Dirac representation via complex joint probabilities \cite{Johansen07,Hofmann12}.
Reviews and additional references on weak measurements can be found in \cite{Resch,APT10,KAN2012}.

The results in \cite{LSPSB2011} demonstrate that the wave function of a pure state can be directly obtained from a single experimental setup, and that one can directly measure both the absolute values and the phases of the coefficients of a pure state in a certain basis. However, the scheme proposed in \cite{LSPSB2011} applies only to pure states, and it is not efficient because most of the data
are discarded due to the post-selection. Here, we
propose tomography schemes for both pure states
and mixed states via weak measurements. Our schemes are more efficient because we do not discard any data. We also show that any (diagonal or off-diagonal) element of the density
state can be directly determined as the average pointer shift in an appropriate weak measurement.

\bigskip

\noindent
{\large \bf Results}

\noindent
{\bf The idea. }
In the scheme proposed in \cite{LSPSB2011}, the wave function to
be measured is a continuous function of the position. The
essence of the scheme is reviewed in finite dimensional Hilbert space as follows \cite{LB12,Salvail12}.
Suppose a quantum system with a $d$-dimensional Hilbert space is in an unknown state
\begin{equation}
\left|\psi\right\rangle =\sum_{i=0}^{d-1}\psi_{i}\left|a_{i}\right\rangle ,
\end{equation}
where the coefficients $\{\psi_{i}\}$ in a certain basis $\{\left|a_{i}\right\rangle \}$ need to be determined.
The scheme for measuring the coefficients
$\{\psi_{i}\}$ consists of a series of weak
measurements of the observables $A_{i}=\left|a_{i}\right\rangle \left\langle a_{i}\right|$,
each followed by the same postselection, i.e., a projection onto the final
state $\left|b_{0}\right\rangle =\frac{1}{\sqrt{d}}\sum_{i=0}^{d-1}\left|a_{i}\right\rangle $.
The weak value of the observable $A_{i}$ is defined as
\begin{equation}
W_{i}=\frac{\left\langle b_{0}|a_{i}\right\rangle \left\langle a_{i}|\psi\right\rangle }{\left\langle b_{0}|\psi\right\rangle }=\frac{1}{\sqrt{d}\left\langle b_{0}|\psi\right\rangle }\psi_{i}. \label{eq:Wi}
\end{equation}
Both the real and the imaginary parts of the weak value $W_i$ have physical meanings and can be determined experimentally because they correspond to the average shifts of the pointer position and the momentum, respectively.
From Eq. (\ref{eq:Wi}), we know that the weak values $W_{i}$ are directly
proportional to the coefficients $\psi_{i}$ that we want to measure and that the
factor $\frac{1}{\sqrt{d}\left\langle b_{0}|\psi\right\rangle }$
can be determined by the normalisation condition (up to an unphysical
overall phase of $\left|\psi\right\rangle $). Therefore, we obtain a direct measurement
of the coefficients $\psi_{i}$, thus a direct measurement of the
wave function $\left|\psi\right\rangle $.

The essential point in this scheme is the choice of the post-selected
state $\left|b_{0}\right\rangle $. Its overlap with each basis state $\left| a_i \right\rangle$
has the same magnitude and phase; therefore,
the factor $K=\psi_{i}/W_{i}=\sqrt{d}\left\langle b_{0}|\psi\right\rangle $
does not depend on $i$ and can be determined by normalisation. Because
the change of the system state due to a weak interaction with the
pointer is negligible, the success probability of post-selection is
given by $P=|\left\langle b_{0}|\psi\right\rangle |^{2}$. Therefore,
only a fraction $P$ of the data is retained, and the remainder is discarded
due to the post-selection. When the dimension $d$ is large, as in the case of a continuous wave function, the majority of the data is discarded.

\medskip
\noindent
{\bf State tomography of a pure state.}
The part of the data that corresponds to the failure of the post-selection
can be retained. We replace the final post-selection by
a complete projective measurement onto the basis states $\left\{ \left|b_{j}\right\rangle \right\} $.
The inner products $\beta_{ji}=\left\langle b_{j}|a_{i}\right\rangle $
are fixed when the two sets of the basis states are chosen. We organise
the data according to the final state: if the final state is $\left|b_{j}\right\rangle$, then
the weak value of $A_{i}=\left|a_{i}\right\rangle \left\langle a_{i}\right|$
is given by
\begin{equation}
W_{ji}=\frac{\left\langle b_{j}|a_{i}\right\rangle \left\langle a_{i}|\psi\right\rangle }{\left\langle b_{j}|\psi\right\rangle }=\frac{\left\langle b_{j}|a_{i}\right\rangle }{\left\langle b_{j}|\psi\right\rangle }\psi_{i}=\frac{\beta_{ji}}{\left\langle b_{j}|\psi\right\rangle }\psi_{i}.
\end{equation}
Therefore, for each fixed $j$, the weak values of different $A_{i}$
give the relative ratios of the coefficients $\psi_{i}$:
\begin{equation}
\psi_{i}^{(j)}=W_{ji}\frac{\left\langle b_{j}|\psi\right\rangle }{\left\langle b_{j}|a_{i}\right\rangle }=\frac{W_{ji}}{\beta_{ji}}\left\langle b_{j}|\psi\right\rangle .\label{eq:pure2nd}
\end{equation}
Because $\left\langle b_{j}|\psi\right\rangle $ does not depend on $i$,
the coefficients $\psi_{i}$ are directly proportional to $\frac{W_{ji}}{\beta_{ji}}$,
where the weak values $W_{ji}$ are read-outs directly from the
experiments and the constants $\beta_{ji}=\left\langle b_{j}|a_{i}\right\rangle $
are fixed when the two sets of basis states are chosen. The super index
$(j)$ in $\psi_{i}^{(j)}$ denotes the coefficients $\psi_{i}$ that
we obtain from the $j$th subset of the data. Ideally, $\psi_{i}^{(j)}$
should not depend on $j$; however, they may depend on $j$ due to imperfections
and noise in the experiments. The magnitude of the factor $\left\langle b_{j}|\psi\right\rangle =e^{i\varphi_{j}}|\left\langle b_{j}|\psi\right\rangle |$
can be removed by normalisation:
\begin{equation}
\psi_{i}^{(j)}=e^{i\varphi_{j}}\frac{W_{ji}}{\beta_{ji}}/\sqrt{\sum_{i'}|\frac{W_{ji'}}{\beta_{ji'}}|^{2}}.\label{eq:pure2nd-1}
\end{equation}
The phase factor $\varphi_{j}$ has no physical meaning because it corresponds
to the overall phase of the pure state we are attempting to measure. This factor can be
removed by a convention. For example, we require $\psi_{0}>0$ (or
the coefficient with the largest magnitude be positive).
Using this approach,
we retain all of the data and separate the data into $d$ subsets according
to the final states $\left\{ \left|b_{j}\right\rangle \right\} $.
For the $j$th subset of the data that corresponds to the fixed final
states $\left|b_{j}\right\rangle $, we have an estimation of the
coefficients $\psi_{i}^{(j)}$ from Eq. (\ref{eq:pure2nd}) or (\ref{eq:pure2nd-1}). A different
subset of the data gives a different estimation for the set of coefficients
$\{\psi_{i}\}$. The differences between the estimations
indicate the amount of error or noise in the experiment.

When the basis $\left\{ \left|b_{j}\right\rangle \right\} $ is mutually
unbiased to the basis $\left\{ \left|a_{i}\right\rangle \right\} $,
i.e., $\beta_{ji}=\left\langle b_{j}|a_{i}\right\rangle =\frac{e^{i\phi_{ji}}}{\sqrt{d}}$,
(\ref{eq:pure2nd-1}) can be simplified to
\begin{equation}
\psi_{i}^{(j)}=e^{i\varphi_{j}}e^{-i\phi_{ji}}W_{ji}/\sqrt{\sum_{i'}|W_{ji'}|^{2}}.
\end{equation}
The original scheme proposed in \cite{LSPSB2011} corresponds to the
case in which $\left|b_{0}\right\rangle =\frac{1}{\sqrt{d}}\sum_{i=0}^{d-1}\left|a_{i}\right\rangle $,
and only the data corresponding to the successful post-selection of
$\left|b_{0}\right\rangle $ are retained. Here, we retain all of the data, and each
subset of the weak values corresponding to a fixed final state gives
an estimation of the state to be measured. The probability of obtaining
the final state $\left|b_{j}\right\rangle $ is given by $|\left\langle b_{j}|\psi\right\rangle |^{2}$
because the change of the system state during the weak interaction is negligible.
When the state $\left|\psi\right\rangle $ to be measured is almost
orthogonal to a certain final state $\left|b_{j}\right\rangle $,
then the relative frequency to obtain the final state $\left|b_{j}\right\rangle $ in the post-selection
is small. However, the corresponding weak value has a large magnitude, which could result in a large signal-to-noise ratio when the dominant noise is due to systematic error
or imperfections.

In the above scheme, weak measurements for a complete set
of projective operators $\{A_{i}=\left|a_{i}\right\rangle \left\langle a_{i}\right|\}$ are needed.
For the tomography of a pure-state wave function $\left| \psi \right\rangle$, weak measurements
of a single observable instead of a set of observables are sufficient. This single observable can be chosen as follows (an alternative choice is presented in the Methods).
We consider a pure state $\left| \varphi \right\rangle$ such that its overlap with each postselected state is nonzero, i.e., $\left\langle b_j | \varphi \right\rangle \neq 0$ for each $j$.  We perform a weak measurement of $P_\varphi =\left|\varphi \right\rangle \left\langle \varphi \right|$.
When the post-selected state is $\left| b_j \right\rangle$, the weak value of $P_{\varphi}$ is defined as
\begin{equation}
W_{j}=\frac{\left\langle b_{j}\right|P_{\varphi}\left|\psi\right\rangle }{\left\langle b_{j}|\psi\right\rangle }
=\frac{\left\langle b_{j}|\varphi \right\rangle  \left\langle \varphi|\psi\right\rangle }{\left\langle b_{j}|\psi\right\rangle } .
\label{wvsingleprojectiveobervable}
\end{equation}
Therefore,
\begin{equation}
\left\langle b_j | \psi \right\rangle = \frac{\left\langle b_j | \varphi \right\rangle}{W_j} \left\langle \varphi | \psi \right\rangle = \eta_j  \left\langle \varphi | \psi \right\rangle .
\label{wvsingleprojectiveobervable11}
\end{equation}
Here, $\eta_j = \frac{\left\langle b_j | \varphi \right\rangle}{W_j} $ is completely determined from the experimental data and our choice because
$W_j$ is directly obtained from the experiment and both $\left| \varphi\right\rangle$
and $\{ \left| b_j \right\rangle\}$ are fixed by our choice.
From Eq. (\ref{wvsingleprojectiveobervable11}), we can reconstruct the state $\left| \psi \right\rangle$:
\begin{equation}
\left| \psi \right\rangle = \sum_j K \eta_j  \left| b_j  \right\rangle ,
\label{wvsingleprojectiveobervable22}
\end{equation}
where $K$ is determined by the normalisation condition up to a phase.

\medskip

\noindent
{\bf State tomography of a mixed state.}
To this point, we have only considered tomography schemes for a pure state.
If the state we like to measure is a general mixed state $\rho$,
we also have a tomography scheme via weak measurements. For this case, we perform
weak measurements for a complete set of projective operators $A_{i}=\left|a_{i}\right\rangle \left\langle a_{i}\right|$, followed by
a final projective measurement onto the basis $\left\{ \left|b_{j}\right\rangle \right\} $.
For the initial state $\rho$ and the final post-selected state $\left|b_{j}\right\rangle $, the average shifts of the pointer position and the momentum in a weak measurement of $A_{i}$ correspond to the real and imaginary parts of the weak value \cite{Wiseman02, WL2011}
\begin{equation}
W_{ji}=\frac{tr\{(\left|b_{j}\right\rangle \left\langle b_{j}\right|)(\left|a_{i}\right\rangle \left\langle a_{i}\right|)\rho\}}{tr(\left|b_{j}\right\rangle \left\langle b_{j}\right|\rho)}=\frac{\left\langle b_{j}|a_{i}\right\rangle \left\langle a_{i}\right|\rho\left|b_{j}\right\rangle }{\left\langle b_{j}\right|\rho\left|b_{j}\right\rangle } .
\label{eq:Wjirho}
\end{equation}
From Eq. (\ref{eq:Wjirho}), one can express the matrix elements of $\rho$
either by the basis $\left\{ \left|a_{i}\right\rangle \right\} $
\begin{equation}
\left\langle a_{i}\right|\rho\left|a_{j}\right\rangle =\sum_{k}\left\langle b_{k}\right|\rho\left|b_{k}\right\rangle \frac{\left\langle b_{k}|a_{j}\right\rangle }{\left\langle b_{k}|a_{i}\right\rangle }W_{ki}=\sum_{k}P_{k}\frac{\beta_{kj}}{\beta_{ki}}W_{ki}\label{eq:rhoabasis}
\end{equation}
or by the basis $\left\{ \left|b_{j}\right\rangle \right\} $
\begin{equation}
\left\langle b_{i}\right|\rho\left|b_{j}\right\rangle =\left\langle b_{j}\right|\rho\left|b_{j}\right\rangle \sum_{k}W_{jk}\frac{\left\langle b_{i}|a_{k}\right\rangle }{\left\langle b_{j}|a_{k}\right\rangle }=P_{j}\sum_{k}W_{jk}\frac{\beta_{ik}}{\beta_{jk}} ,
\label{eq:rhoBbasis}
\end{equation}
where the probability $P_{j}=\left\langle b_{j}\right|\rho\left|b_{j}\right\rangle$  corresponds to the frequency of
obtaining the final state $\left|b_{j}\right\rangle $ in the experiment.
The weak values $W_{ji}$ and the probabilities
$P_{j}$ are directly accessible from the experiment. The relationships
in Eqs. (\ref{eq:rhoabasis}) and (\ref{eq:rhoBbasis}) show that the
matrix elements of the mixed state can be written as linear summations
of the weak values that are directly accessible from the experiment.

\medskip

\noindent
{\bf Partial state tomography.}
Using the idea of weak measurements, we can also selectively measure
some of the matrix elements (diagonal and off-diagonal terms) of a
state directly. This is especially useful when we are only interested
in one or a few matrix elements of the state and we do not need to
perform state tomography for the entire density matrix. Suppose we are
interested in measuring a particular matrix element $\left\langle a\right|\rho\left|b\right\rangle $.
We consider two different cases according to whether $\left| a \right\rangle$ and $\left| b \right\rangle$ are orthogonal.
For the first case, suppose $\left\langle b|a\right\rangle \neq0$. Here, we can simply
perform a weak measurement of $\left|a\right\rangle \left\langle a\right|$,
with a post-selection onto the state $\left|b\right\rangle $. The
weak value is defined as
\begin{equation}
W=\frac{\left\langle b|a\right\rangle \left\langle a\right|\rho\left|b\right\rangle }{\left\langle b\right|\rho\left|b\right\rangle } .
\end{equation}
Thus,
\begin{equation}
\left\langle a\right|\rho\left|b\right\rangle =\frac{\left\langle b\right|\rho\left|b\right\rangle }{\left\langle b|a\right\rangle }W . \label{onematrixelement}
\end{equation}
$W$ is the weak value, and $\left\langle b\right|\rho\left|b\right\rangle $
is the probability of success for post-selection. Both values are directly
obtained from the experiment. $\left\langle b|a\right\rangle $
is only a fixed factor. Therefore, $\left\langle a\right|\rho\left|b\right\rangle $
is determined by Eq. (\ref{onematrixelement}).  For the second case, suppose that $\left\langle b|a\right\rangle =0$.
We then perform a weak measurement of the observable $C=\left|c\right\rangle \left\langle c\right|$ with
$\left|c\right\rangle =\frac{1}{\sqrt{2}}(\left|a\right\rangle +\left|b\right\rangle )$
and a follow-up projective measurement onto a basis
including both $\left|a\right\rangle $ and $\left|b\right\rangle $
as the basis states. We must only retain the data when the final
state is either $\left|a\right\rangle $ or $\left|b\right\rangle $.
If the final state is $\left|a\right\rangle $, then the weak value of
$C$ is given by
\begin{equation}
W=\frac{\left\langle a|c\right\rangle \left\langle c\right|\rho\left|a\right\rangle }{\left\langle a\right|\rho\left|a\right\rangle }=\frac{1}{2}(1+\frac{\left\langle b\right|\rho\left|a\right\rangle }{\left\langle a\right|\rho\left|a\right\rangle }) .
\end{equation}
If the final state is $\left|b\right\rangle $, then the weak value
of $C$ is given by
\begin{equation}
W'=\frac{\left\langle b|c\right\rangle \left\langle c\right|\rho\left|b\right\rangle }{\left\langle b\right|\rho\left|b\right\rangle }=\frac{1}{2}(1+\frac{\left\langle a\right|\rho\left|b\right\rangle }{\left\langle b\right|\rho\left|b\right\rangle }).
\end{equation}
We have
\begin{eqnarray}
\left\langle b\right|\rho\left|a\right\rangle &=& \left\langle a\right|\rho\left|a\right\rangle (2W-1)\label{eq:matrixelementbrhoa} ,  \\
\left\langle a\right|\rho\left|b\right\rangle &=&\left\langle b\right|\rho\left|b\right\rangle (2W'-1)\label{eq:matrixelementsanotherarhob} .
\end{eqnarray}
The post-selection probabilities $\left\langle a\right|\rho\left|a\right\rangle $ and
$\left\langle b\right|\rho\left|b\right\rangle $ and the weak values
$W$ and $W'$ are directly accessible from the experiment; therefore,
we can determine the matrix element $\left\langle a\right|\rho\left|b\right\rangle$ from either Eq. (\ref{eq:matrixelementbrhoa})
or (\ref{eq:matrixelementsanotherarhob}), which should agree with each other. Any discrepancies
between them could be used as indicators of the noise and imperfections
in the experiment.

\bigskip

\noindent
{\bf Experimental scheme.} Here, we discuss how our schemes can be realised experimentally. For the weak measurement of an observable $A$,
the interaction between the measuring device and the system
is generally modelled by a Hamiltonian $\mathcal{H}=gA\otimes p\delta(t-t_{0})$,
where $p$ denotes the momentum of the pointer (measuring device).
Because the interaction in a weak measurement does not change the state
of the system significantly, weak measurements of several observables
(whether they are commutative or not) can be performed simultaneously.
The interaction of the simultaneous weak measurements of a set of observables
$\{A_{i}\}$ can be introduced by the Hamiltonian:
\begin{equation}
\mathcal{H}=\sum_{i}g_{i}A_{i}\otimes p_{i}\delta(t-t_{0}) . \label{eq:hamiltonian}
\end{equation}
Here, $p_{i}$ denotes the momentum of
the $i$th pointer that is coupled to the observable $A_{i}$.  $\{A_{i}\}$ is an arbitrary set of observables of the system. They could
be a set of commuting observables (for example, $A_{i}=\left|a_{i}\right\rangle \left\langle a_{i}\right|$)
or even a set of complementary observables.  Suppose
the initial state of the system is $\rho_{s}$, and the initial state
of the pointers is $\rho_{d}=\rho_{1}\otimes\rho_{2}\otimes\cdots$,
where $\rho_{i}$ denotes the initial state of the $i$th pointer.
Suppose the post-selection of the system is a projection $\Pi_{j}$
(for example, $\Pi_{j}=\left|b_{j}\right\rangle \left\langle b_{j}\right|$).
Then the average position shift $\delta q_{i}$ and the average momentum shift $\delta p_{i}$
of the $i$th pointer are given by (see the derivation in the Methods)
\begin{eqnarray}
\delta q_{i}&=&g_{i}ReW_{ji} \label{eq:deltaqi}    \\
\delta p_{i}&=& 2g_{i}ImW_{ji}(\Delta p_{i})^{2} .
\label{eq:deltapi}
\end{eqnarray}
The weak value $W_{ji}$ of $A_i$, for a general initial state $\rho_s$ and a general postselection $\Pi_{j}$, is given by \cite{WZ2012}
\begin{equation}
W_{ji}=\frac{Tr(\Pi_{j}A_{i}\rho_{s})}{Tr(\Pi_{j}\rho_{s})} .
\end{equation}
The shifts of different pointers can be read out simultaneously, while
either the position shift or the momentum shift is
recorded at each time for each pointer. With the same Hamiltonian, we can obtain all of the weak values $W_{ji}$.
From these weak values, we can obtain all of the elements of the density matrix $\rho_{s}$
according to the state tomography strategy proposed in this article.

\medskip

\noindent {\bf Example of application.}
The state tomography schemes proposed here can also be used to detect a tiny parameter encoded in a quantum state.
For example, to detect a tiny phase difference $\vartheta$ picked up by two orthogonal states of a qubit (a photon) that passes through a certain medium (a birefringent crystal), we could prepare a qubit in the initial state $\frac{1}{\sqrt{2}} \left( \left| 0 \right\rangle +  \left| 1 \right\rangle \right)$, and seek to determine the tiny phase $\vartheta$ encoded in the outgoing state
$\frac{1}{\sqrt{2}} \left( \left| 0 \right\rangle + e^{i \vartheta} \left| 1 \right\rangle \right)$.
To determine $\vartheta$, we can perform a weak measurement of the observable $\left| 1 \right\rangle \left\langle 1 \right|$ by introducing an interaction $\mathcal{H}=g \left| 1 \right\rangle \left\langle 1 \right| \otimes p \delta(t-t_{0})$ between the qubit system and a measuring device and a follow-up post-selection of the qubit system onto the state $\frac{1}{\sqrt{2}} \left( \left| 0 \right\rangle - \left| 1 \right\rangle \right)$.
The weak value is $W=1-i\frac{1}{\vartheta}$, and the average shift of the pointer momentum is given by
$\delta p  = - 2g (\Delta p)^{2} \frac{1}{\vartheta}$.  Although the phase $\vartheta$ is small, the average shift of the pointer momentum is not necessarily small, and this has certain advantages over the quantum-entanglement based scheme in Ramsey interferometry \cite{GLM2011}, which relies on the use of GHZ states of many qubits, which are an expensive resource.

\bigskip

\noindent
{\large \bf Discussion}

\noindent
In this article, we have proposed several schemes for state tomography
via weak measurements. An advantage of our schemes is that
we need to measure fewer observables compared with the standard method of
state tomography based on ideal measurements. For example, a simple
standard state tomography requires ideal measurements in at least $d+1$ different
bases; therefore, one needs to use at least $d+1$ different experimental setups.
Here, our tomography strategy requires weak interaction in one
basis ($\left\lbrace \left|a_{i}\right\rangle \right\rbrace $) and
post-selection in another basis ($\left\lbrace \left|b_{j}\right\rangle \right\rbrace $), and
the same interaction Hamiltonian is used for all data collection.

Compared with the scheme in \cite{LSPSB2011}, our strategy applies not only to pure states but also to mixed states.
Our strategy is more efficient because we retain all of the data and thus reduces the number of experimental runs needed.
Compared with the schemes in \cite{LB12} in which a density matrix is directly measured via sequential weak measurements of pairs or triple products of observables, our schemes are based on a single-time weak measurement followed by a strong complete projective measurement. Our interaction Hamiltonian in Eq. (\ref{eq:hamiltonian}) is straightforward and differs significantly from those in the schemes of \cite{LB12}. In the previous schemes \cite{LSPSB2011,LB12}, the observables $A_i=\left| a_{i}\right\rangle \left\langle a_{i}\right|$ require separate weak measurements at each time for a particular $i$. In our schemes, the set of observables $\{ A_i \}$ can be measured simultaneously via the interaction introduced by the simple Hamiltonian in Eq. (\ref{eq:hamiltonian}).

Interestingly, any matrix element of a density
state is directly accessible from a suitable weak measurement.
Ideal measurements give probabilities and expectation values, which are only linear combinations of the matrix elements of the density operator.
Based on all of these facts, it is natural to draw the conclusion:
{\bf what is and can be really measured in a general (ideal or weak) measurement is the density matrix (with all its elements), and only the density matrix!}

\bigskip
\noindent
{\large \bf Acknowledgements}

\noindent
The author likes to thank Marek \.Zukowski for valuable discussions, and wishes to acknowledge support from the National Natural Science Foundation of China (Grants No. 11075148 and No. 11275181), the Chinese Academy of Sciences, and the National Fundamental Research Program.

\bigskip

\noindent
{\large \bf Methods}

\noindent
{\bf An alternative scheme for the tomography of a pure state via the weak measurement of a single observable.}
In the main text, we see that it is sufficient to perform a weak measurement
of a single observable for the tomography of a pure-state wave function $\left| \psi \right\rangle$.
This single observable could be chosen alternatively as follows.
We introduce a single observable
\begin{equation}
A=\sum_{i} \lambda_{i}\left|a_{i}\right\rangle \left\langle a_{i}\right| ,
\end{equation}
where the eigenvalues $\lambda_{i}$ are not degenerated, i.e., $\lambda_{i}\neq \lambda_{j}$
if $i\neq j$. We simply perform a weak measurement of the observable
$A$ and retain all of the data corresponding to different final states $\left| b_j\right\rangle$.
When the final state is $\left|b_{j}\right\rangle $, the weak value
of $A$ is given by
\begin{equation}
W_{j}=\frac{\left\langle b_{j}\right|A\left|\psi\right\rangle }{\left\langle b_{j}|\psi\right\rangle }=\frac{\sum_{i}\left\langle b_{j}|a_{i}\right\rangle \lambda_{i}\left\langle a_{i}|\psi\right\rangle }{\left\langle b_{j}|\psi\right\rangle }=\frac{\sum_{i}\beta_{ji}\lambda_{i}\psi_{i}}{\sum_{i}\beta_{ji}\psi_{i}}\label{eq:wjsingleobervable}
\end{equation}
From Eq. (\ref{eq:wjsingleobervable}) we have
\begin{equation}
\sum_{i}(\beta_{ji}\lambda_{i}-W_{j}\beta_{ji})\psi_{i}=0.\label{eq:wjsinglecoefficienteq}
\end{equation}
We define a unitary matrix $\beta$ with the matrix elements given by the
coefficients $\beta_{ji}$ and introduce the diagonal matrices $\lambda=diag\{\lambda_{0},\lambda_{1},\cdots,\lambda_{d-1}\}$,
$w=diag\{w_{0},w_{1},\cdots,w_{d-1}\}$ and the vector $\overrightarrow{\psi}=(\psi_{0},\psi_{1},\cdots,\psi_{d-1})^{T}$,
where $T$ denotes the transpose. Then, Eq. (\ref{eq:wjsinglecoefficienteq})
can be written as
\begin{equation}
(\beta \lambda-w\beta)\overrightarrow{\psi}=0.
\end{equation}
With the notation $M=\beta \lambda-w\beta$, we have $M\overrightarrow{\psi}=0$.
Thus, the state tomography in this case is to find the vector $\overrightarrow{\psi}$
corresponding to the zero eigenvalue of $M$. The matrix $M$ depends
on $w$, which is determined by the weak values that are directly read out from the experiments. Due
to noise and imperfections in the experiments, matrix $M$
may not have a zero eigenvalue. Instead, multiplying $M\overrightarrow{\psi}=0$
by $M^{\dagger}$ from the left, we have
\begin{equation}
(M^{\dagger}M)\overrightarrow{\psi}=0.
\end{equation}
Therefore, the vector $\overrightarrow{\psi}$ is the eigenvector
in the kernel space of the non-negative matrix $M^{\dagger}M$. The
vector $\overrightarrow{\psi}$ is uniquely determined if $M^{\dagger}M$
has a one-dimensional kernel space. It is not determined, and another
set of measurements must be performed if $M^{\dagger}M$ has a kernel space of more than one dimension.
However, it is also possible that the matrix $M^{\dagger}M$ determined from the experiment may not
have a kernel space. In such a case, we choose the most likely one, i.e., we choose the vector $\overrightarrow{\psi}$
as the eigenvector corresponding to the smallest eigenvalue of $M^{\dagger}M$.
The value of the smallest eigenvalue of $M^{\dagger}M$ indicates the
amount of noise and imperfection in the experiment.

\medskip
\noindent
{\bf Derivation of the position shift and the momentum shift.}
Here, we derive the position shift and the momentum shift in (\ref{eq:deltaqi})
and (\ref{eq:deltapi}), respectively. The time evolution operator corresponding
to the interaction Hamiltonian in (\ref{eq:hamiltonian}) is given
by $U=e^{-i\sum_{i}g_{i}A_{i}\otimes p_{i}}$. If the system is successfully
post-selected by $\Pi_{j}$, which occurs with a probability (to the
first order of $g_{i}$) of
\begin{eqnarray}
P_{j} & = & Tr((\Pi_{j}\otimes I)U\rho_{s}\otimes\rho_{d}U^{\dagger}) \nonumber \\
 & = & Tr(\Pi_{j}\rho_{s})(1+2\sum_{i}g_{i}ImW_{ji}\left\langle p_{i}\right\rangle )
\end{eqnarray}
the final (unnormalised) state of the pointers is given by (to the
first order of $g_{i}$)
\begin{eqnarray}
\rho'_{d} & = & Tr_{s}((\Pi_{j}\otimes I)U\rho_{s}\otimes\rho_{d}U^{\dagger})\\
 & = & Tr(\Pi_{j}\rho_{s})(\rho_{d}-i\sum_{i}g_{i}(W_{ji}p_{i}\rho_{d}-W_{ji}^{*}\rho_{d}p_{i})).
\end{eqnarray}
Here, $Tr_{s}$ ($Tr$) denotes the trace over the system (whole)
Hilbert space. The average position shift of the $i$th pointer, conditional
on the successful post-selction by $\Pi_{j}$, is given by $\delta q_{i}=\frac{Tr(q_{i}\rho'_{d})}{Tr(\rho'_{d})}-\left\langle q_{i}\right\rangle $.
A straightforward calculation to the first order of $g_{i}$ gives
\begin{equation}
\delta q_{i}= g_{i}ReW_{ji}+g_{i}ImW_{ji}\left\langle \left\lbrace p_{i}-\left\langle p_{i}\right\rangle ,\ q_{i}-\left\langle q_{i}\right\rangle \right\rbrace \right\rangle . \label{eq:deltaqimethod}
\end{equation}
When the initial states of the pointers are well-behaved states, such as Gaussian states,
the second term on the right-hand side of (\ref{eq:deltaqimethod}) vanishes, and we immediately have
(\ref{eq:deltaqi}). Similarly the average momentum shift of the $i$th
pointer, conditional on the successful post-selection by $\Pi_{j}$,
is given by $\delta p_{i}=\frac{Tr(p_{i}\rho'_{d})}{Tr(\rho'_{d})}-\left\langle p_{i}\right\rangle $,
which yields (\ref{eq:deltapi}) to the first order of $g_i$.

\end{document}